\documentclass[pra]{revtex4}
\usepackage{amssymb} \usepackage{graphicx}

\begin{document}
 \title{Lie algebra of conformal Killing-Yano forms}

\author{\"Umit Ertem}
 \email{umitertemm@gmail.com}
\address{Department of Physics,
Ankara University, Faculty of Sciences, 06100, Tando\u gan-Ankara,
Turkey\\}
 
\begin{abstract}

We provide a generalization of the Lie algebra of conformal Killing vector fields to conformal Killing-Yano forms. A new Lie bracket for conformal Killing-Yano forms that corresponds to slightly modified Schouten-Nijenhuis bracket of differential forms is proposed. We show that conformal Killing-Yano forms satisfy a graded Lie algebra in constant curvature manifolds. It is also proven that normal conformal Killing-Yano forms in Einstein manifolds also satisfy a graded Lie algebra. The constructed graded Lie algebras reduce to the graded Lie algebra of Killing-Yano forms and the Lie algebras of conformal Killing and Killing vector fields in special cases.

\end{abstract}

\maketitle

\section{Introduction}

Killing vector fields generate the isometries of a given manifold. Their flows preserve the metric defined on it and the number of Killing vectors corresponds to the symmetries of the manifold. The manifolds that have maximal number of Killing vectors are called maximally symmetric. Under the Lie bracket of vector fields, Killing vectors satisfy a Lie algebra that corresponds to the Lie algebra of the isometry group of the manifold. Similarly, the vector fields that preserve a conformal class of metrics are called conformal Killing vector fields. They also form a Lie algebra with respect to the Lie bracket of vector fields which is called the conformal algebra. Killing vector fields correspond to a subset of conformal Killing vector fields. Besides the symmetric tensor generalizations called Killing tensors and conformal Killing tensors, they also have antisymmetric generalizations called Killing-Yano (KY) forms and conformal Killing-Yano (CKY) forms respectively \cite{Yano, Tachibana Kashiwada}. There are various aspects of them that have importance in physical problems. For example, the first integrals of the geodesic motion of a particle can be constructed from KY forms \cite{Hughston Penrose Sommers Walker}. They are also used in the construction of symmetry operators of the massive Dirac equation in curved backgrounds \cite{Benn Kress, Acik Ertem Onder Vercin1}. Similarly CKY forms are used for the symmetry operator construction of massless Dirac equation \cite{Benn Charlton}. Conserved gravitational currents which are constructed from KY forms and curvature characteristics of the manifold can be related to $p$-brane charges \cite{Kastor Traschen, Acik Ertem Onder Vercin2, Ertem Acik}. Moreover, the $p$-form Dirac currents of geometric Killing spinors correspond to KY forms and they can be used in the construction of superalgebras in supersymmetric field theories \cite{Acik Ertem, OFarrill HackettJones Moutsopoulos Simon, Ertem}. Similary, the $p$-form Dirac currents of twistor spinors are CKY forms and can be related to conformal superalgebras in supergravity backgrounds \cite{Acik Ertem, de Medeiros Hollands}.

Lie algebra of Killing vectors can be generalized to KY forms in constant curvature manifolds. They satisfy a graded Lie algebra structure under Schouten-Nijenhuis (SN) bracket in those cases \cite{Kastor Ray Traschen}. This corresponds to the extended symmetry algebra of the background. Different algebra structures for closed CKY forms that correspond to the Hodge duals of KY forms also exist in the literature which are applicable to rotating black hole spacetimes \cite{Krtous Kubiznak Page Frolov, Cariglia Krtous Kubiznak}. However, there is no known result for the extension of the Lie algebra of conformal Killing vectors to all CKY forms. In this paper, we propose a Lie bracket for CKY forms and show that they satisfy a graded Lie algebra or a Lie superalgebra under this bracket in constant curvature manifolds. For the subset of KY forms, it reduces to the graded Lie algebra of KY forms with respect to the SN bracket. Moreover, a subset of CKY forms which are called normal CKY forms also satisfy a graded Lie algebra in Einstein manifolds with respect to the defined bracket. This provides a first step for the construction of conformal superalgebras that include the graded Lie algebra of CKY forms which are extensions of the conformal superalgebras constructed out of conformal Killing vectors.

The paper is organized as follows. In Section 2, we consider the CKY equation and its integrability conditions in terms of the curvature characteristics of the manifold. We also define the subset of normal CKY forms and the relevant integrability conditions. In Section 3, we propose a Lie bracket for CKY forms and show that the bracket of two normal CKY forms is also a normal CKY form in Einstein manifolds and it is a CKY form for all CKY forms in constant curvature manifolds. We also show the graded Lie bracket and Lie superalgebra properties of the new bracket. Section 4 concludes the paper.

\section{CKY Forms and Integrability Conditions}

A $p$-form $\omega$ is a CKY $p$-form in an $n$-dimensional manifold, if it satisfies the following equation which is the antisymmetric generalization of conformal Killing equation to higher-degree differential forms
\begin{equation}
\nabla_{X}\omega=\frac{1}{p+1}i_{X}d\omega-\frac{1}{n-p+1}\widetilde{X}\wedge\delta\omega
\end{equation}
where $X$ is any vector field, $\widetilde{X}$ its metric dual, $\nabla$ is the Levi-Civita connection, $i_X$ is the interior derivative or contraction with respect to $X$, $d$ and $\delta$ are exterior derivative and co-derivative operators respectively. Co-closed CKY forms, namely $\delta\omega=0$, correspond to KY forms.

CKY equation is invariant under Hodge star operation which is denoted by *. If $\omega$ is a CKY $p$-form, then $*\omega$ is a CKY $(n-p)$-form in an $n$-dimensional manifold and we have the following equation
\begin{equation}
\nabla_{X}*\omega=\frac{1}{n-p+1}i_{X}d*\omega-\frac{1}{p+1}\widetilde{X}\wedge\delta*\omega.
\end{equation}
This can be seen from the properties of * and the equality $\nabla_X*=*\nabla_X$. CKY equation is also covariant under conformal rescalings of the metric. If $\omega$ is a CKY $p$-form in a background with metric $g$, then for a background with metric $\widehat{g}=e^{2\lambda}g$, the $p$-form $\widehat{\omega}=e^{(p+1)\lambda}\omega$ is again a CKY $p$-form where $\lambda$ is a function. This can be proven directly from (1) by using the following relations between operators defined for the metric $g$ and conformally rescaled metric $\widehat{g}$
\begin{eqnarray}
\widehat{\nabla}_X\omega&=&\nabla_X\omega-pX(\lambda)\omega-d\lambda\wedge i_X\omega+\widetilde{X}\wedge i_{\widetilde{d\lambda}}\omega\nonumber\\
\widehat{d}\omega&=&d\omega\\
\widehat{\delta}\omega&=&\delta\omega-(n-2p)i_{\widetilde{d\lambda}}\omega\nonumber.
\end{eqnarray}

Integrability conditions of CKY forms can be obtained by taking second derivatives of the CKY equation \cite{Semmelmann}. The curvature operator on a manifold is defined as follows
\begin{equation}
R(X,Y)=[\nabla_X,\nabla_Y]-\nabla_{[X,Y]}
\end{equation}
where $X$ and $Y$ are arbitrary vector fields. For the ease of calculations, we will choose a frame basis ${X_a}$ in normal coordinates, so the connection coefficients will be zero and we have $\nabla_{X_a}X_b=0=[X_a,X_b]$. By taking the second covariant derivative of (1), we obtain
\begin{eqnarray}
\nabla_{X_a}\nabla_{X_b}\omega&=&\frac{1}{p+1}\nabla_{X_a}i_{X_b}d\omega-\frac{1}{n-p+1}\nabla_{X_a}\left(e_b\wedge\delta\omega\right)\nonumber\\
&=&\frac{1}{p+1}i_{X_b}\nabla_{X_a}d\omega-\frac{1}{n-p+1}e_b\wedge\nabla_{X_a}\delta\omega
\end{eqnarray}
where ${e_b}$ is coframe basis with the property $e_b(X_a)=\delta_{ba}$ and we have used the relation $[i_X, \nabla_Y]=i_{\nabla_XY}$. By using (5), the action of the curvature operator on a CKY $p$-form can be obtained in the following form
\begin{eqnarray}
R(X_a,X_b)\omega&=&\nabla_{X_a}\nabla_{X_b}\omega-\nabla_{X_b}\nabla_{X_a}\omega\nonumber\\
&=&\frac{1}{p+1}\left(i_{X_b}\nabla_{X_a}-i_{X_a}\nabla_{X_b}\right)d\omega-\frac{1}{n-p+1}\left(e_b\wedge\nabla_{X_a}-e_a\wedge\nabla_{X_b}\right)\delta\omega.
\end{eqnarray}
By taking the wedge product with $e^a$, (6) transforms into
\begin{equation}
e^a\wedge R(X_a,X_b)\omega=-\frac{p}{p+1}\nabla_{X_b}d\omega+\frac{1}{n-p+1}e_b\wedge d\delta\omega
\end{equation}
where we have used the relations $e^a\wedge i_{X_a}\omega=p\omega$ and $e^a\wedge\nabla_{X_a}\omega=d\omega$ with the definition of the Lie derivative ${\cal{L}}_X=i_Xd+di_X$ and the properties $[{\cal{L}}_X,d]=0$ and $d^2=0$. Since the action of the curvature operator on a $p$-form can be written in terms of curvature 2-forms $R_{ab}$ as $e^a\wedge R(X_a,X_b)\omega=-R_{ab}\wedge i_{X^a}\omega$ (see \cite{Acik Ertem Onder Vercin2}), the first integrability condition of the CKY equation is obtained as follows
\begin{equation}
\nabla_{X_a}d\omega=\frac{p+1}{p(n-p+1)}e_ a\wedge d\delta\omega+\frac{p+1}{p}R_{ba}\wedge i_{X^b}\omega.
\end{equation}

On the other hand, by taking the interior product of (6) with $i_{X^a}$ we obtain
\begin{equation}
i_{X^a}R(X_a,X_b)\omega=\frac{1}{p+1}i_{X_b}\delta d\omega+\frac{n-p}{n-p+1}\nabla_{X_b}\delta\omega
\end{equation}
where we have used the relations $i_{X_a}i_{X_b}=-i_{X_b}i_{X_a}$ and $i_{X^a}\nabla_{X_a}=-\delta$ with $\delta^2=0$. Since the action of the curvature operator on $\omega$ can be written in terms of curvature 2-forms $R_{ab}$ and Ricci 1-forms $P_b=i_{X^a}R_{ab}$ as $i_{X^a}R(X_a,X_b)\omega=i_{X_c}P_b\wedge i_{X^c}\omega+i_{X_c}R_{ab}\wedge i_{X^a}i_{X^c}\omega$, the second integrability condition of the CKY equation is found as
\begin{equation}
\nabla_{X_a}\delta\omega=-\frac{n-p+1}{(p+1)(n-p)}i_{X_a}\delta d\omega+\frac{n-p+1}{n-p}\left(i_{X_b}P_a\wedge i_{X^b}\omega+i_{X_b}R_{ca}\wedge i_{X^c}i_{X^b}\omega\right).
\end{equation}
In fact, by using the definitions $\delta d\omega=-i_{X^a}\nabla_{X_a}d\omega$ and $d\delta\omega=e^a\wedge\nabla_{X_a}\delta\omega$, the integrability conditions (8) and (10) can be combined into the following equality
\begin{equation}
\frac{p}{p+1}\delta d\omega+\frac{n-p}{n-p+1}d\delta\omega=P_a\wedge i_{X^a}\omega+R_{ab}\wedge i_{X^a}i_{X^b}\omega
\end{equation}
where we have used the identities $i_{X_a}P_b=i_{X_b}P_a$ and $i_{X_a}R_{bc}+i_{X_b}R_{ca}+i_{X_c}R_{ab}=0$ which are relevant for zero torsion.

\subsection{Normal CKY forms}

A subset of CKY forms can be defined in terms of the following 1-form which plays an important role in conformal geometry and corresponds to the metric dual of the so-called Schouten tensor
\begin{equation}
K_a=\frac{1}{n-2}\left(\frac{\cal{R}}{2(n-1)}e_a-P_a\right)
\end{equation}
where ${\cal{R}}=i_{X^a}P_a$ is the curvature scalar. A CKY $p$-form $\omega$ is called a normal CKY $p$-form, if it satisfies the following integrability conditions besides the CKY equation \cite{Leitner, Lischewski}
\begin{equation}
\nabla_{X_a}d\omega=\frac{p+1}{p(n-p+1)}e_a\wedge d\delta\omega+2(p+1)K_a\wedge\omega
\end{equation}
\begin{equation}
\nabla_{X_a}\delta\omega=-\frac{n-p+1}{(p+1)(n-p)}i_{X_a}\delta d\omega-2(n-p+1)i_{X^b}K_a\wedge i_{X_b}\omega.
\end{equation}
These correspond to special cases of (8) and (10). For normal CKY forms, (11) is written as
\begin{equation}
\frac{p}{p+1}\delta d\omega+\frac{n-p}{n-p+1}d\delta\omega=-2(n-p)K_a\wedge i_{X^a}\omega.
\end{equation}
Moreover, in constant curvature manifols, (13) and (14) are equal to (8) and (10) respectively. This can be seen as follows. For a constant curvature manifold, the curvature 2-forms are written as $R_{ab}=\frac{\cal{R}}{n(n-1)}e_a\wedge e_b$ and the Ricci 1-forms are $P_a=\frac{\cal{R}}{n}e_a$. So, the curvature parts of (8) and (10) are written as
\begin{eqnarray}
\frac{p+1}{p}R_{ba}\wedge i_{X^b}\omega&=&-\frac{p+1}{n(n-1)}{\cal{R}}e_a\wedge\omega\nonumber\\
\frac{n-p+1}{n-p}\left(i_{X_b}P_a\wedge i_{X^b}\omega+i_{X_b}R_{ca}\wedge i_{X^c}i_{X^b}\omega\right)&=&\frac{n-p+1}{n(n-1)}{\cal{R}}i_{X_a}\omega
\end{eqnarray}
and the right hand sides of (16) are also equivalent to the curvature terms in (13) and (14)
\begin{eqnarray}
2(p+1)K_a\wedge\omega&=&-\frac{p+1}{n(n-1)}{\cal{R}}e_a\wedge\omega\nonumber\\
-2(n-p+1)i_{X^b}K_a\wedge i_{X_b}\omega&=&\frac{n-p+1}{n(n-1)}{\cal{R}}i_{X_a}\omega.
\end{eqnarray}
This means that in constant curvature manifolds, all CKY forms are normal CKY forms. So, in that case the integrability conditions are written in the following form
\begin{equation}
\nabla_{X_a}d\omega=\frac{p+1}{p(n-p+1)}e_a\wedge d\delta\omega-\frac{p+1}{n(n-1)}{\cal{R}}e_a\wedge\omega
\end{equation}
\begin{equation}
\nabla_{X_a}\delta\omega=-\frac{n-p+1}{(p+1)(n-p)}i_{X_a}\delta d\omega+\frac{n-p+1}{n(n-1)}{\cal{R}}i_{X_a}\omega
\end{equation}
\begin{equation}
\frac{p}{p+1}\delta d\omega+\frac{n-p}{n-p+1}d\delta\omega=\frac{p(n-p)}{n(n-1)}{\cal{R}}\omega.
\end{equation}

The motivation for defining normal CKY forms is the generalization of the properties of CKY forms in constant curvature or conformally flat manifolds to the general cases. Integrability conditions of CKY forms in constant curvature case are more well-behaved in the calculations than the general case and normal CKY forms share this good properties in Einstein manifolds and other cases. Indeed, normal CKY forms correspond to the parallel sections of conformal Cartan conection and they are important for the calculation of the holonomy of the Cartan connection. So, they can be used in the classification of the conformal geometries \cite{Leitner}. There are some special physical backgrounds that have normal CKY forms besides the constant curvature spacetimes like de Sitter ($dS$) or anti-de Sitter ($AdS$) spacetimes in which there are maximal number of CKY forms that are also normal CKY forms. For example, gravitational instantons constructed from Eguchi-Hanson metric have normal CKY forms that are composed of twistor spinors in the background \cite{Kuhnel Rademacher}. Generally, the squaring map of twistor spinors give normal CKY forms. Moreover, Fefferman spaces in Lorentzian geometry and pp-wave spacetimes also have normal CKY forms \cite{Leitner}.

\section{A Lie bracket for CKY Forms}

Killing vector fields and conformal Killing vector fields constitute a Lie algebra structure under the ordinary Lie bracket of vector fields. It is defined for the vector fields $X$ and $Y$ as follows
\begin{eqnarray}
[X,Y]=\nabla_XY-\nabla_YX.\nonumber
\end{eqnarray}
The natural generalization of the Lie bracket of vector fields to higher-degree differential forms is the SN bracket and it is defined for a $p$-form $\alpha$ and a $q$-form $\beta$ as
\begin{equation}
[\alpha, \beta]_{SN}=i_{X_a}\alpha\wedge\nabla_{X^a}\beta+(-1)^p\nabla_{X_a}\alpha\wedge i_{X^a}\beta.
\end{equation}
The SN bracket has the graded Lie bracket properties and it is known that KY forms satisfy a graded Lie algebra under SN bracket in constant curvature spacetimes \cite{Kastor Ray Traschen}. However, there is no known such a result about the Lie algebra structure of antisymmetric generalizations of conformal Killing vectors that are CKY forms. Although it can be thought that the SN bracket can provide a graded Lie algebra structure for CKY forms as a generalization of ordinary Lie bracket of conformal Killing vector fields, this is not the case even in constant curvature manifolds. Instead of this, we propose another generalization of the Lie bracket of vector fields which is slightly different from the SN bracket and provide a graded Lie algebra structure for CKY forms in some special cases. The bracket that we propose for a CKY $p$-form $\omega_1$ and a CKY $q$-form $\omega_2$ is the following
\begin{eqnarray}
[\omega_1, \omega_2]_{CKY}&=&\frac{1}{q+1}i_{X_a}\omega_1\wedge i_{X^a}d\omega_2+\frac{(-1)^p}{p+1}i_{X_a}d\omega_1\wedge i_{X^a}\omega_2\nonumber\\
&&+\frac{(-1)^p}{n-q+1}\omega_1\wedge\delta\omega_2+\frac{1}{n-p+1}\delta\omega_1\wedge\omega_2.
\end{eqnarray}
This gives a ($p+q-1$)-form and for $p=q=1$, it reduces to the Lie bracket of vector fields. For the subset of KY forms which correspond to the co-closed CKY forms that have the property $\delta\omega=0$, this bracket reduces to the SN bracket of KY forms. However, for general CKY forms, it differs from the SN bracket in terms of the coefficients of the third and fourth terms in (22). We will show that the CKY bracket defined in (22) is a graded Lie bracket for normal CKY forms in Einstein manifolds and for all CKY forms in constant curvature manifolds. Namely, they satisfy the following equality in those cases
\begin{equation}
\nabla_{X_a}[\omega_1, \omega_2]_{CKY}=\frac{1}{p+q}i_{X_a}d[\omega_1, \omega_2]_{CKY}-\frac{1}{n-p-q+2}e_a\wedge\delta[\omega_1, \omega_2]_{CKY}.
\end{equation}

To prove (23), we first calculate the covariant derivative of CKY bracket defined in (22)
\begin{eqnarray}
\nabla_{X_a}[\omega_1, \omega_2]_{CKY}&=&\frac{1}{q+1}i_{X_b}\nabla_{X_a}\omega_1\wedge i_{X^b}d\omega_2+\frac{1}{q+1}i_{X_b}\omega_1\wedge i_{X^b}\nabla_{X_a}d\omega_2\nonumber\\
&&+\frac{(-1)^p}{p+1}i_{X_b}\nabla_{X_a}d\omega_1\wedge i_{X^b}\omega_2+\frac{(-1)^p}{p+1}i_{X_b}d\omega_1\wedge i_{X^b}\nabla_{X_a}\omega_2\nonumber\\
&&+\frac{(-1)^p}{n-q+1}\nabla_{X_a}\omega_1\wedge\delta\omega_2+\frac{(-1)^p}{n-q+1}\omega_1\wedge\nabla_{X_a}\delta\omega_2\nonumber\\
&&+\frac{1}{n-p+1}\nabla_{X_a}\delta\omega_1\wedge\omega_2+\frac{1}{n-p+1}\delta\omega_1\wedge\nabla_{X_a}\omega_2
\end{eqnarray}
where we have used $[i_{X_b}, \nabla_{X_a}]=0$ for normal coordinates. We will use the CKY equation (1) and the integrability conditions (8) and (10) by defining the following curvature quantities of a $p$-form $\omega_1$ and similarly of a $q$-form $\omega_2$ for the ease of calculations
\begin{eqnarray}
\alpha_pM_{1a}&:=&\frac{p+1}{p}R_{ba}\wedge i_{X^b}\omega_1\\
\alpha_qM_{2a}&:=&\frac{q+1}{q}R_{ba}\wedge i_{X^b}\omega_2\nonumber\\
\gamma_pN_{1a}&:=&\frac{n-p+1}{n-p}\left(i_{X_b}P_a\wedge i_{X^b}\omega_1+i_{X_b}R_{ca}\wedge i_{X^c}i_{X^b}\omega_1\right)\\
\gamma_qN_{2a}&:=&\frac{n-q+1}{n-q}\left(i_{X_b}P_a\wedge i_{X^b}\omega_2+i_{X_b}R_{ca}\wedge i_{X^c}i_{X^b}\omega_2\right).\nonumber
\end{eqnarray}
So, we obtain
\begin{eqnarray}
\nabla_{X_a}[\omega_1, \omega_2]_{CKY}&=&\frac{1}{(p+1)(q+1)}i_{X_b}i_{X_a}d\omega_1\wedge i_{X^b}d\omega_2+\frac{(-1)^p}{(p+1)(q+1)}i_{X_b}d\omega_1\wedge i_{X^b}i_{X_a}d\omega_2\nonumber\\
&&+\frac{1}{q(n-q+1)}i_{X_a}\omega_1\wedge d\delta\omega_2+\frac{(-1)^p}{p(n-p+1)}d\delta\omega_1\wedge i_{X_a}\omega_2\nonumber\\
&&-\frac{1}{(p+1)(n-p)}i_{X_a}\delta d\omega_1\wedge\omega_2-\frac{(-1)^p}{(q+1)(n-q)}\omega_1\wedge i_{X_a}\delta d\omega_2\nonumber\\
&&+\frac{1}{(n-p+1)(q+1)}e_a\wedge i_{X_b}\delta\omega_1\wedge i_{X^b}d\omega_2+\frac{(-1)^p}{(p+1)(n-q+1)}i_{X_b}d\omega_1\wedge e_a\wedge i_{X^b}\delta\omega_2\nonumber\\
&&-\frac{1}{q(n-q+1)}i_{X_b}\omega_1\wedge e_a\wedge i_{X^b}d\delta\omega_2-\frac{(-1)^p}{p(n-p+1)}e_a\wedge i_{X_b}d\delta\omega_1\wedge i_{X^b}\omega_2\nonumber\\
&&+\frac{\alpha_q}{q+1}i_{X_b}\omega_1\wedge i_{X^b}M_{2a}+\frac{(-1)^p\alpha_p}{p+1}i_{X_b}M_{1a}\wedge i_{X^b}\omega_2\nonumber\\
&&+\frac{\gamma_p}{n-p+1}N_{1a}\wedge\omega_2+\frac{(-1)^p\gamma_q}{n-q+1}\omega_1\wedge N_{2a}
\end{eqnarray}
By taking the wedge product of (27) with $e_a$ from the left and then applying the interior product operation, we arrive at the first term of the right hand side of (23) as follows
\begin{eqnarray}
\frac{1}{p+q}i_{X_a}d[\omega_1, \omega_2]_{CKY}&=&\frac{1}{(p+1)(q+1)}i_{X_b}i_{X_a}d\omega_1\wedge i_{X^b}d\omega_2+\frac{(-1)^p}{(p+1)(q+1)}i_{X_b}d\omega_1\wedge i_{X^b}i_{X_a}d\omega_2\nonumber\\
&&+\frac{1}{q(n-q+1)}i_{X_a}\omega_1\wedge d\delta\omega_2-\frac{(-1)^p}{(q+1)(n-q)}\omega_1\wedge i_{X_a}\delta d\omega_2\nonumber\\
&&-\frac{1}{(p+1)(n-p)}i_{X_a}\delta d\omega_1\wedge\omega_2+\frac{(-1)^p}{p(n-p+1)}d\delta\omega_1\wedge i_{X_a}\omega_2\nonumber\\
&&+\frac{\alpha_q}{(q+1)(p+q)}\left(i_{X_b}\omega_1\wedge i_{X^b}M_{2a}-e^c\wedge i_{X_a}\left(i_{X_b}\omega_1\wedge i_{X^b}M_{2c}\right)\right)\nonumber\\
&&+\frac{(-1)^p\alpha_p}{(p+1)(p+q)}\left(i_{X_b}M_{1a}\wedge i_{X^b}\omega_2-e^c\wedge i_{X_a}\left(i_{X_b}M_{1c}\wedge i_{X^b}\omega_2\right)\right)\nonumber\\
&&+\frac{\alpha_qq}{(q+1)(n-q)(p+q)}i_{X_a}\omega_1\wedge i_{X^b}M_{2b}+\frac{(-1)^p\alpha_pp}{(p+1)(n-p)(p+q)}i_{X^b}M_{1b}\wedge i_{X_a}\omega_2\nonumber\\
&&+\frac{(-1)^p\gamma_q}{q(n-q+1)}\left(\omega_1\wedge N_{2a}-e^b\wedge\omega_1\wedge i_{X_a}N_{2b}\right)\nonumber\\
&&+\frac{\gamma_p}{p(n-p+1)}\left(N_{1a}\wedge\omega_2-e^b\wedge i_{X_a}N_{1b}\wedge\omega_2\right)\nonumber\\
&&-\frac{(-1)^p\gamma_q}{(n-q+1)(p+q)}e^b\wedge i_{X_a}\omega_1\wedge N_{2b}+\frac{(-1)^p\gamma_p}{(n-p+1)(p+q)}e^b\wedge N_{1b}\wedge i_{X_a}\omega_2
\end{eqnarray}
where we have used the integrability condition (11). Similarly, applying the interior product operation to (27) and then taking the wedge product with $e_a$ from the left gives the second term on the right hand side of (23)
\begin{eqnarray}
-\frac{1}{n-p-q+2}e_a\wedge\delta[\omega_1,\omega_2]_{CKY}&=&\frac{1}{(n-p+1)(q+1)}e_a\wedge i_{X_b}\delta\omega_1\wedge i_{X^b}d\omega_2+\frac{(-1)^p}{q(n-q+1)}e_a\wedge i_{X_b}\omega_1\wedge i_{X^b}d\delta\omega_2\nonumber\\
&&+\frac{1}{(p+1)(n-q+1)}e_a\wedge i_{X_b}d\omega_1\wedge i_{X^b}\delta\omega_2-\frac{(-1)^p}{p(n-p+1)}e_a\wedge i_{X_b}d\delta\omega_1\wedge i_{X^b}\omega_2\nonumber\\
&&-\frac{(-1)^p\alpha_p(n-p+1)}{(p+1)(n-p)(n-p-q+2)}e_a\wedge i_{X_b}i_{X^c}M_{1c}\wedge i_{X^b}\omega_2\nonumber\\
&&+\frac{(-1)^p\alpha_q(n-q+1)}{(q+1)(n-q)(n-p-q+2)}e_a\wedge i_{X_b}\omega_1\wedge i_{X^b}i_{X^c}M_{2c}\nonumber\\
&&-\frac{1}{(n-p-q+2)}\left(\frac{\alpha_p}{p+1}e_a\wedge i_{X_b}M_{1c}\wedge i_{X^c}i_{X^b}\omega_2-\frac{\alpha_q}{q+1}e_a\wedge i_{X^c}i_{X^b}\omega_1\wedge i_{X^b}M_{2c}\right)\nonumber\\
&&+\frac{\gamma_q}{(n-q+1)(n-p-q+2)}\left(e_a\wedge\omega_1\wedge i_{X^b}N_{2b}+(-1)^pe_a\wedge i_{X^b}\omega_1\wedge N_{2b}\right)\nonumber\\
&&+\frac{\gamma_p}{(n-p+1)(n-p-q+2)}\left(e_a\wedge i_{X^b}N_{1b}-(-1)^pe_a\wedge N_{1b}\wedge i_{X^b}\omega_2\right).
\end{eqnarray}
Now, we have all the ingredients to check the Lie bracket property of the CKY bracket defined in (22). First, we start with flat  manifolds in which we have all the curvature terms equal to zero $M_a=0=N_a$. It can easily be seen that the non-curvature terms in (27) are equivalent to the non-curvature terms in the sum of (28) and (29). Hence, we prove that the CKY bracket of CKY $p$-form $\omega_1$ and CKY $q$-form $\omega_2$ is again a CKY $(p+q-1)$-form in flat manifolds. On the other hand, the curvature terms in (27), (28) and (29) are not equal to each other in manifolds of arbitrary curvature. However, we will see in the next subsection that the Lie bracket property can still be satisfied in some manifolds of special curvature such as constant curvature and Einstein manifolds.

\subsection{Constant curvature manifolds}

We first consider the manifolds with constant curvature in which the curvature 2-forms are written as $R_{ab}=\frac{\cal{R}}{n(n-1)}e_a\wedge e_b$. In this case, the curvature terms defined in (25) and (26) transforms into
\begin{eqnarray}
\alpha_p M_{1a}&:=&-\frac{p+1}{n(n-1)}{\cal{R}}e_a\wedge\omega_1\\
\alpha_q M_{2a}&:=&-\frac{q+1}{n(n-1)}{\cal{R}}e_a\wedge\omega_2\nonumber\\
\gamma_p N_{1a}&:=&\frac{n-p+1}{n(n-1)}{\cal{R}} i_{X_a}\omega_1\\
\gamma_q N_{2a}&:=&\frac{n-q+1}{n(n-1)}{\cal{R}} i_{X_a}\omega_2.\nonumber
\end{eqnarray}
From these definitions, it can easily be seen that the curvature terms in (27) vanish. Moreover, the sum of the curvature terms in (28) and (29) is exactly equal to zero. So, this proves that the CKY bracket of two CKY forms in constant curvature manifolds is again a CKY form. The importance of this result is in the fact that the maximally symmetric spacetimes such as $dS$ and $AdS$ backgrounds are constant curvature manifolds and have maximal number of CKY forms. This provides a way to extend the Lie algebra of conformal Killing vector fields to CKY forms and define an extended conformal algebra in maximally symmetric spacetimes of various dimensions.

\subsection{Einstein manifolds}

The second case that we consider is the normal CKY forms in Einstein manifolds. For the normal CKY forms, the curvature terms defined in (25) and (26) are written as
\begin{eqnarray}
\alpha_p M_{1a}&:=&2(p+1)K_a\wedge\omega_1\\
\alpha_q M_{2a}&:=&2(q+1)K_a\wedge\omega_2\nonumber\\
\gamma_p N_{1a}&:=&-2(n-p+1)i_{X^b}K_a\wedge i_{X_b}\omega_1\\
\gamma_q N_{2a}&:=&-2(n-q+1)i_{X^b}K_a\wedge i_{X_b}\omega_2.\nonumber
\end{eqnarray}
So, the curvature terms in (27) are found to be zero
\begin{eqnarray}
&&2i_{X_b}\omega_1\wedge\left(i_{X^b}K_a\wedge\omega_2-K_a\wedge i_{X^b}\omega_2\right)+(-1)^p2\left(i_{X_b}K_a\wedge\omega_1-K_a\wedge i_{X_b}\omega_1\right)\wedge i_{X^b}\omega_2\nonumber\\
&&-(-1)^p
2\omega_1\wedge i_{X^b}K_a\wedge i_{X_b}\omega_2-2i_{X^b}K_a\wedge i_{X_b}\omega_1\wedge\omega_2\nonumber\\
&&=0.
\end{eqnarray}
We need to prove that the curvature terms in the sum of (28) and (29) also vanish. In Einstein manifolds, the Ricci 1-forms satisfy the relation $P_a=\frac{\cal{R}}{n}e_a$. Hence, we have
\begin{eqnarray}
K_a=-\frac{\cal{R}}{2n(n-1)}e_a
\end{eqnarray}
and
\begin{eqnarray}
\alpha_p M_{1a}&:=&-\frac{p+1}{n(n-1)}{\cal{R}}e_a\wedge\omega_1\\
\alpha_q M_{2a}&:=&-\frac{q+1}{n(n-1)}{\cal{R}}e_a\wedge\omega_2\nonumber\\
\gamma_p N_{1a}&:=&\frac{n-p+1}{n(n-1)}{\cal{R}} i_{X_a}\omega_1\\
\gamma_q N_{2a}&:=&\frac{n-q+1}{n(n-1)}{\cal{R}} i_{X_a}\omega_2.\nonumber
\end{eqnarray}
These are exactly the same as in the case of (30) and (31). So, the curvature terms in the sum of (28) and (29) are exactly equal to zero. This means that the CKY bracket of two normal CKY forms in Einstein manifolds is again a CKY form. Moreover, it is a normal CKY form since the curvature characteristics defined in the definition of the normal CKY forms in (13) and (14) reduces to the curvature characteristics of ordinary CKY forms in Einstein manifolds. To prove the Lie algebra structure of CKY forms, we will consider the Lie bracket properties of the CKY bracket in the next subsection.

\subsection{Lie bracket properties of the CKY bracket}

The CKY bracket defined in (22) satisfy the graded Lie bracket properties. Those are the following (skew)-symmetry condition for a $p$-form $\alpha$ and a $q$-form $\beta$
\begin{equation}
[\alpha, \beta]_{CKY}=(-1)^{pq}[\beta, \alpha]_{CKY}
\end{equation}
and the graded Jacobi identity
\begin{equation}
(-1)^{p(r+1)}[\alpha, [\beta, \gamma]_{CKY}]_{CKY}+(-1)^{q(p+1)}[\beta, [\gamma, \alpha]_{CKY}]_{CKY}+(-1)^{r(q+1)}[\gamma, [\alpha, \beta]_{CKY}]_{CKY}=0
\end{equation}
where $\gamma$ is an $r$-form. In fact, one can see from (38) that for odd forms the CKY bracket is a skew-symmetric bracket while for even forms it is a symmetric bracket. This means that CKY forms in constant curvature manifolds and normal CKY forms in Einstein manifolds satisfy a Lie superalgebra structure with respect to the CKY bracket.

A Lie superalgebra $\mathfrak{g}=\mathfrak{g}_0\oplus\mathfrak{g}_1$ consists of an even part $\mathfrak{g}_0$ which is a Lie algebra and an odd part $\mathfrak{g}_1$ which is a vector space that is acted on by $\mathfrak{g}_0$. There is a bilinear multiplication $[.,.]$ on the superalgebra that satisfies the following (skew)-supersymmetry and super-Jacobi identities
\begin{eqnarray}
[a,b]&=&-(-1)^{|a||b|}[b,a]\nonumber\\
\left[a,[b,c]\right]&=&[[a, b], c]+(-1)^{|a||b|}[b, [a, c]]
\end{eqnarray}
where $a, b, c$ are elements of $\mathfrak{g}$ and $|a|$ denotes the degree of $a$ which corresponds to 0 or 1 depending on $a$ is in $\mathfrak{g}_0$ or $\mathfrak{g}_1$, respectively. For the case of CKY superalgebra, $\mathfrak{g}_0$ corresponds to the Lie algebra of odd CKY forms and $\mathfrak{g}_1$ corresponds to the space of even CKY forms. The properties (38) and (39) of the CKY bracket corresponds to the Lie superalgebra properties in (40).

For the subset of KY forms, namely CKY forms $\omega$ that satisfy $\delta\omega=0$, the CKY superalgebra reduces to the Lie superalgebra of KY forms with respect to the SN bracket. For $p=q=1$, they correspond to the Lie algebra of conformal Killing vector fields and Killing vector fields respectively.

Graded Lie algebra of CKY forms can be used in the extended conformal superalgebras of supersymmetric field theories in different curved backgrounds. Conformal Killing vector fields and twistor spinors are main tools in the construction of conformal symmetry superalgebras in various supergravity backgrounds. By using the Lie algebra structure of conformal Killing vector fields and adding an extra R-symmetry to the theory, the conformal superalgebra gain a Lie superalgebra structure \cite{de Medeiros Hollands, Leitner}. Lie algebra of conformal Killing vectors correspond to the even part of the superalgebra and twistor spinors constitute the odd part of it. So, the Lie algebra of odd CKY forms in maximally symmetric spacetimes is a natural candidate for the extended conformal superalgebras of supersymmetric field theories in relevant maximally symmetric backgrounds. The similar extension can be done for conformal superalgebras in Einstein manifolds from the Lie algebra of normal odd CKY forms. These investigations can give way to obtain different supersymmetric field theories in various supergravity backgrounds.

\section{Conclusion}

Lie algebra of Killing vector fields can be generalized to KY forms in constant curvature manifolds by the SN bracket of differential forms. A generalization of the Lie algebra of conformal Killing vector fields is provided by defining a new bracket for CKY forms. Although it is similar to the SN bracket, it differs from it in the coefficients of some terms. By computing the covariant derivative of the CKY bracket and compraring it with the exterior and co-derivatives of the bracket, it is shown that CKY forms constitute a graded Lie algebra in constant curvature manifolds. A similar graded Lie algebra structure can also be constructed for normal CKY forms in Einstein manifolds. In fact, they are Lie superalgebras and the even part of it corresponds to the Lie algebra of odd CKY forms while the odd part of it is equal to the space of even CKY forms.

The construction of the graded Lie algebra of CKY forms can be related to the investigation of extended superalgebras in supergravity theories. Killing superalgebras in supergravity backgrounds are constructed out of Killing spinors and Killing vector fields. The even part of the superalgebra corresponds to the Lie algebra of Killing vector fields. Similarly, conformal superalgebras in supergravity backgrounds can be constructed out of twistor spinors and conformal Killing vector fields by using the Lie algebra of conformal Killing vector fields. They are used for obtaining results about the classification problem of supergravity backgrounds. The construction of the graded Lie algebras of KY forms and CKY forms can be used to extend the Killing superalgebras to include higher-degree differential forms. It is known that the higher-degree Dirac currents of Killing spinors correspond to KY forms while for twistor spinors they are equal to CKY forms \cite{Acik Ertem}. So, the odd parts of the superalgebras can generate the even parts of them. An investigation for the extensions of Killing superalgebras in $AdS_5$ background is done in \cite{Ertem}. Similar analysis is under investigation for conformal superalgebras by using the construction of the graded Lie algebra of CKY forms.

\begin{acknowledgments}

The author thanks School of Mathematics of The University of Edinburgh for the kind hospitality and providing a fruitful scientific atmosphere during his stay in Edinburgh where this work started.

\end{acknowledgments}


\end{document}